\input harvmac
%
%
%
%
\font\ticp=cmcsc10

\def\hf{{1\over 2}}
\def\calo{{\cal O}}

\def\cald{{\cal D}}

\def\vhat{{\hat v}}

\def\calA{{\cal A}}
\def\calL{{\cal L}}

\def\mthsu{\mathsurround=0pt  }
\def\leftrightarrowfill{$\mthsu \mathord\leftarrow\mkern-6mu\cleaders
  \hbox{$\mkern-2mu \mathord- \mkern-2mu$}\hfill
  \mkern-6mu\mathord\rightarrow$}
 \def\overleftrightarrow#1{\vbox{\ialign{##\crcr\leftrightarrowfill\crcr\noalign{\kern-1pt\nointerlineskip}$\hfil\displaystyle{#1}\hfil$\crcr}}}
\overfullrule=0pt
%
%
\lref\Bousrev{
  R.~Bousso,
  ``The holographic principle,''
  Rev.\ Mod.\ Phys.\  {\bf 74}, 825 (2002)
  [arXiv:hep-th/0203101].
}
\lref\ACVone{
  D.~Amati, M.~Ciafaloni and G.~Veneziano,
  ``Superstring Collisions At Planckian Energies,''
  Phys.\ Lett.\ B {\bf 197}, 81 (1987).
}
\lref\ACVtwo{
  D.~Amati, M.~Ciafaloni and G.~Veneziano,
  ``Classical And Quantum Gravity Effects From Planckian Energy Superstring
  Collisions,''
  Int.\ J.\ Mod.\ Phys.\ A {\bf 3}, 1615 (1988).
}
\lref\ACVthree{
  D.~Amati, M.~Ciafaloni and G.~Veneziano,
  ``Higher Order Gravitational Deflection And Soft Bremsstrahlung In Planckian
  Energy Superstring Collisions,''
  Nucl.\ Phys.\ B {\bf 347}, 550 (1990).
}
\lref\ACVfour{
  D.~Amati, M.~Ciafaloni and G.~Veneziano,
  ``Effective action and all order gravitational eikonal at Planckian
  energies,''
  Nucl.\ Phys.\ B {\bf 403}, 707 (1993).
}
\lref\Dono{
  J.~F.~Donoghue,
  ``General relativity as an effective field theory: The leading quantum
  corrections,''
  Phys.\ Rev.\ D {\bf 50}, 3874 (1994)
  [arXiv:gr-qc/9405057]\semi
  ``Introduction to the Effective Field Theory Description of Gravity,''
  arXiv:gr-qc/9512024.
}
\lref\tHooholo{
  G.~'t Hooft,
  ``Dimensional reduction in quantum gravity,''
  arXiv:gr-qc/9310026.
  }
  \lref\MuSo{
  I.~J.~Muzinich and M.~Soldate,
  ``High-Energy Unitarity Of Gravitation And Strings,''
  Phys.\ Rev.\ D {\bf 37}, 359 (1988).
}
\lref\Sold{
  M.~Soldate,
  ``Partial Wave Unitarity And Closed String Amplitudes,''
  Phys.\ Lett.\ B {\bf 186}, 321 (1987).
}
\lref\GRW{
  G.~F.~Giudice, R.~Rattazzi and J.~D.~Wells,
  ``Transplanckian collisions at the LHC and beyond,''
  Nucl.\ Phys.\ B {\bf 630}, 293 (2002)
  [arXiv:hep-ph/0112161].
}
\lref\Hsu{
  S.~D.~H.~Hsu,
  ``Quantum production of black holes,''
  Phys.\ Lett.\ B {\bf 555}, 92 (2003)
  [arXiv:hep-ph/0203154].
}
\lref\KaOr{
  D.~Kabat and M.~Ortiz,
  ``Eikonal quantum gravity and Planckian scattering,''
  Nucl.\ Phys.\ B {\bf 388}, 570 (1992)
  [arXiv:hep-th/9203082].
}
\lref\Sussholo{
  L.~Susskind,
  ``The World as a hologram,''
  J.\ Math.\ Phys.\  {\bf 36}, 6377 (1995)
  [arXiv:hep-th/9409089].
}
\lref\tHooftRE{
  G.~'t Hooft,
  ``On The Quantum Structure Of A Black Hole,''
  Nucl.\ Phys.\ B {\bf 256}, 727 (1985).
}
\lref\tHooftFR{
  G.~'t Hooft,
  ``The Black Hole Interpretation Of String Theory,''
  Nucl.\ Phys.\ B {\bf 335}, 138 (1990).
}
\lref\Mald{
  J.~M.~Maldacena,
  ``The large N limit of superconformal field theories and supergravity,''
  Adv.\ Theor.\ Math.\ Phys.\  {\bf 2}, 231 (1998)
  [Int.\ J.\ Theor.\ Phys.\  {\bf 38}, 1113 (1999)]
  [arXiv:hep-th/9711200].
}
\lref\SusskindIF{
  L.~Susskind, L.~Thorlacius and J.~Uglum,
  ``The Stretched horizon and black hole complementarity,''
  Phys.\ Rev.\ D {\bf 48}, 3743 (1993)
  [arXiv:hep-th/9306069].
}
\lref\tHooftRB{
  G.~'t Hooft,
  ``The Black hole horizon as a quantum surface,''
  Phys.\ Scripta {\bf T36}, 247 (1991).
}
\lref\Bous{
  R.~Bousso,
  ``A Covariant Entropy Conjecture,''
  JHEP {\bf 9907}, 004 (1999)
  [arXiv:hep-th/9905177]\semi
  ``Holography in general space-times,''
  JHEP {\bf 9906}, 028 (1999)
  [arXiv:hep-th/9906022].
}
\lref\SGinfo{S.~B.~Giddings,
  ``Quantum mechanics of black holes,''
  arXiv:hep-th/9412138.
}
\lref\BHMR{
  S.~B.~Giddings,
  ``Black holes and massive remnants,''
  Phys.\ Rev.\ D {\bf 46}, 1347 (1992)
  [arXiv:hep-th/9203059].
}
\lref\Astroinfo{
  A.~Strominger,
  ``Les Houches lectures on black holes,''
  arXiv:hep-th/9501071.
}
\lref\CaMa{
  C.~G.~.~Callan and J.~M.~Maldacena,
  ``D-brane Approach to Black Hole Quantum Mechanics,''
  Nucl.\ Phys.\ B {\bf 472}, 591 (1996)
  [arXiv:hep-th/9602043].
}
\lref\Hawkunc{
  S.~W.~Hawking,
  ``Breakdown Of Predictability In Gravitational Collapse,''
  Phys.\ Rev.\ D {\bf 14}, 2460 (1976).
}
\lref\StVa{
  A.~Strominger and C.~Vafa,
  ``Microscopic Origin of the Bekenstein-Hawking Entropy,''
  Phys.\ Lett.\ B {\bf 379}, 99 (1996)
  [arXiv:hep-th/9601029].
}
\lref\GMH{
  S.~B.~Giddings, D.~Marolf and J.~B.~Hartle,
  ``Observables in effective gravity,''
  arXiv:hep-th/0512200.
}
\lref\tHooftFN{
G.~'t Hooft,
 ``Fundamental Aspects Of Quantum Theory Related To The Problem Of Quantizing
Black Holes,''
{\it Prepared for Fundamental Aspects of Quantum Theory: Conference on Foundations of Quantum Mechanics to Celebrate 30 Years of the Aharonov-
Bohm Effect (AB Conference), Columbia, South Carolina, 14-16 Dec 1989}.
}
\lref\tHooftUN{
G.~'t Hooft,
``Horizon operator approach to black hole quantization,''
arXiv:gr-qc/9402037.
}
\lref\VerlindeSG{
E.~Verlinde and H.~Verlinde,
``A Unitary S matrix and 2-D black hole formation and evaporation,''
Nucl.\ Phys.\ B {\bf 406}, 43 (1993)
[arXiv:hep-th/9302022].
}
\lref\SchoutensST{
K.~Schoutens, H.~Verlinde and E.~Verlinde,
``Black hole evaporation and quantum gravity,''
arXiv:hep-th/9401081.
}
\lref\KiemIY{
Y.~Kiem, H.~Verlinde and E.~Verlinde,
``Black hole horizons and complementarity,''
Phys.\ Rev.\ D {\bf 52}, 7053 (1995)
[arXiv:hep-th/9502074].
}
\lref\Hawkrec{
  S.~W.~Hawking,
  ``Information loss in black holes,''
  Phys.\ Rev.\ D {\bf 72}, 084013 (2005)
  [arXiv:hep-th/0507171].
}
\lref\LPSTU{
  D.~A.~Lowe, J.~Polchinski, L.~Susskind, L.~Thorlacius and J.~Uglum,
  ``Black hole complementarity versus locality,''
  Phys.\ Rev.\ D {\bf 52}, 6997 (1995)
  [arXiv:hep-th/9506138].
}
\lref\GiLitwo{
  S.~B.~Giddings and M.~Lippert,
  ``The information paradox and the locality bound,''
  Phys.\ Rev.\ D {\bf 69}, 124019 (2004)
  [arXiv:hep-th/0402073].
}
\lref\Jacob{
  T.~Jacobson,
  ``Introduction to quantum fields in curved spacetime and the Hawking
  effect,''
  arXiv:gr-qc/0308048.
}
\lref\GiLione{
  S.~B.~Giddings and M.~Lippert,
  ``Precursors, black holes, and a locality bound,''
  Phys.\ Rev.\ D {\bf 65}, 024006 (2002)
  [arXiv:hep-th/0103231].
}
\lref\AiSe{
  P.~C.~Aichelburg and R.~U.~Sexl,
  ``On The Gravitational Field Of A Massless Particle,''
  Gen.\ Rel.\ Grav.\  {\bf 2}, 303 (1971).
}
\lref\Penrose{R. Penrose, {\sl unpublished} (1974).}
\lref\EaGi{
  D.~M.~Eardley and S.~B.~Giddings,
  ``Classical black hole production in high-energy collisions,''
  Phys.\ Rev.\ D {\bf 66}, 044011 (2002)
  [arXiv:gr-qc/0201034].
}
\lref\Polchrev{
  J.~Polchinski,
  ``String theory and black hole complementarity,''
  arXiv:hep-th/9507094.
}
\lref\BPS{
  T.~Banks, L.~Susskind and M.~E.~Peskin,
  ``Difficulties For The Evolution Of Pure States Into Mixed States,''
  Nucl.\ Phys.\ B {\bf 244}, 125 (1984).
}
\lref\BaFi{T.~Banks and W.~Fischler,
  ``A model for high energy scattering in quantum gravity,''
  arXiv:hep-th/9906038.
}
\lref\GiTh{
  S.~B.~Giddings and S.~D.~Thomas,
  ``High energy colliders as black hole factories: The end of short  distance
  physics,''
  Phys.\ Rev.\ D {\bf 65}, 056010 (2002)
  [arXiv:hep-ph/0106219].
}
\lref\DiLa{
  S.~Dimopoulos and G.~Landsberg,
  ``Black holes at the LHC,''
  Phys.\ Rev.\ Lett.\  {\bf 87}, 161602 (2001)
  [arXiv:hep-ph/0106295].
}
\lref\CGHS{
  C.~G.~Callan, S.~B.~Giddings, J.~A.~Harvey and A.~Strominger,
  ``Evanescent black holes,''
  Phys.\ Rev.\ D {\bf 45}, 1005 (1992)
  [arXiv:hep-th/9111056].
}
\lref\GiNe{
  S.~B.~Giddings and W.~M.~Nelson,
  ``Quantum emission from two-dimensional black holes,''
  Phys.\ Rev.\ D {\bf 46}, 2486 (1992)
  [arXiv:hep-th/9204072].
}
\lref\Hawkevap{
  S.~W.~Hawking,
  ``Particle Creation By Black Holes,''
  Commun.\ Math.\ Phys.\  {\bf 43}, 199 (1975)
  [Erratum-ibid.\  {\bf 46}, 206 (1976)].
}
\lref\Wald{
  R.~M.~Wald,
  ``On Particle Creation By Black Holes,''
  Commun.\ Math.\ Phys.\  {\bf 45}, 9 (1975).
}
\lref\Waldunpub{R.~M.~Wald, unpublished.}
\lref\HoMa{
  G.~T.~Horowitz and J.~Maldacena,
  ``The black hole final state,''
  JHEP {\bf 0402}, 008 (2004)
  [arXiv:hep-th/0310281].
}

\lref\GiRy{
  S.~B.~Giddings and V.~S.~Rychkov,
  ``Black holes from colliding wavepackets,''
  Phys.\ Rev.\ D {\bf 70}, 104026 (2004)
  [arXiv:hep-th/0409131].
}
\lref\LQGST{
  S.~B.~Giddings,
  ``Locality in quantum gravity and string theory,''
  arXiv:hep-th/0604072.
}
\lref\SGnonloc{
  S.~B.~Giddings,
  ``Black hole information, unitarity, and nonlocality,''
  arXiv:hep-th/0605196.
}
\lref\SGinfo{S.~B.~Giddings,
  ``Quantum mechanics of black holes,''
  arXiv:hep-th/9412138\semi
  ``The Black hole information paradox,''
  arXiv:hep-th/9508151.
}
\lref\Astroinfo{
  A.~Strominger,
  ``Les Houches lectures on black holes,''
  arXiv:hep-th/9501071.
}
\lref\Vene{
  G.~Veneziano,
  ``A Stringy Nature Needs Just Two Constants,''
  Europhys.\ Lett.\  {\bf 2}, 199 (1986); talk presented at the Italian Physical Society Meeting, Naples, 1987.
}
\lref\Gross{D.~J.~Gross,
  ``Superstrings And Unification,'' PUPT-1108
{\it Plenary Session talk given at 24th Int. Conf. on High Energy Physics, Munich, West Germany, Aug 4-10, 1988}.
}
\lref\GrMe{
  D.~J.~Gross and P.~F.~Mende,
  ``String Theory Beyond The Planck Scale,''
  Nucl.\ Phys.\ B {\bf 303}, 407 (1988).
}
\lref\Vaid{P.C. Vaidya, Proc. Indian Acad. Sci {\bf A33}, 264 (1951).}
\lref\tHooftTQ{
  G.~'t Hooft,
  ``The scattering matrix approach for the quantum black hole: An overview,''
  Int.\ J.\ Mod.\ Phys.\ A {\bf 11}, 4623 (1996)
  [arXiv:gr-qc/9607022].
}
\lref\KiemIY{
  Y.~Kiem, H.~L.~Verlinde and E.~P.~Verlinde,
  ``Black hole horizons and complementarity,''
  Phys.\ Rev.\ D {\bf 52}, 7053 (1995)
  [arXiv:hep-th/9502074].
}
\lref\ArcioniWC{
  G.~Arcioni,
  ``On 't Hooft's S-matrix ansatz for quantum black holes,''
  JHEP {\bf 0410}, 032 (2004)
  [arXiv:hep-th/0408005].
}
\Title{\vbox{\baselineskip12pt
\hbox{hep-th/0606146}
}}
{\vbox{\centerline{(Non)perturbative gravity, nonlocality, and nice slices}
}}
\centerline{{\ticp Steven B. Giddings}\footnote{$^\star$}{Email address: giddings@physics.ucsb.edu} 
}
\centerline{ \sl Department of Physics}
\centerline{\sl University of California}
\centerline{\sl Santa Barbara, CA 93106-9530}
\bigskip
\centerline{\bf Abstract}

Perturbative dynamics of gravity is investigated for high energy scattering and in black hole backgrounds.  In the latter case, a straightforward perturbative analysis fails, in a close parallel to the failure of the former when the impact parameter reaches the Schwarzschild radius.  This suggests a flaw in a semiclassical description of physics on spatial slices that intersect both outgoing Hawking radiation and matter that has carried information into a black hole; such slices are instrumental in a general argument for black hole information loss.  This indicates a possible role for the proposal that nonperturbative gravitational physics is intrinsically nonlocal.

\Date{}

\newsec{Introduction}

One of the greatest problems in today's physics is understanding dynamics when both quantum-mechanical and gravitational effects are relevant.  While some progress has been made through perturbative and semiclassical methods, as well as string theory, much is lacking, particularly in the strong curvature regime which is believed to arise in ultra-planckian scattering, the interior of black holes, and the early stages of the Universe.   

As experiment is not yet a particularly good guide for these regimes, one should actively exploit any clues to the dynamics.  One that seems particularly prominent is the black hole information paradox, which arose from Hawking's discovery that black holes evaporate\refs{\Hawkevap} and his subsequent argument that they destroy quantum information\refs{\Hawkunc}.  (For reviews of the paradox, see \refs{\SGinfo,\Astroinfo}.)  In particular, one should consider the possibility that this paradox serves as a guidepost to new physics that is as fundamentally important as  the instability of matter in classical physics was towards quantum mechanics.  

In an attempt to avoid the unacceptable features of unitarity violation, a viewpoint has emerged that physics is in some way fundamentally nonlocal; early advocates of this include \refs{\BHMR,\tHooholo,\Sussholo}, and there has been much study of ideas surrounding the holographic proposal that the number of degrees of freedom inside a black hole is given by the Bekenstein-Hawking entropy.  However, what has so far been missing is a more fine-grain characterization of the mechanism for such possible nonlocality, as well as a description of in what regimes it supplants  local field theory, and how  it evades Hawking's semiclassical arguments for information destruction.  

In particular note that there should be something like a correspondence principle for nonlocal physics.  Specifically, if there  is some fundamental nonlocal physics, then it should reduce to familiar local quantum field theory plus perturbative gravity in regimes so far accessible to experiment.  Conversely, one should attempt to parameterize the regimes where such local physics would not be a good approximation to its more fundamental progenitor.  Comparing with quantum mechanics, one seeks a parameterization analogous to, for example, the uncertainty principle, which tells us that classical phase space gives way to quantum wave functions when we attempt to probe it on scales such that $\Delta x\Delta p\sim 1$.  

One can attempt to explore this question in the context of string theory, which many believe to be a consistent quantum theory of gravity.  Indeed, there were early suggestions\refs{\Vene,\Gross} of a  string uncertainty principle of the form 
\eqn\genunc{\Delta x \roughly> {1\over \Delta p} + l_{st}^2 \Delta p\ .}
The second term can be heuristically thought of as arising from production of long strings at high energies; effects of such dynamics were considered in a black hole background  in \refs{\LPSTU}.

More generally, in the context of a $D$-dimensional gravitational theory that may or may not be string theory, a different proposal\refs{\GiLione\GiLitwo\LQGST-\SGnonloc}  is that of a  gravitational nonlocality principle, stating that local quantum physics fails when describing two modes with (approximate) positions and momenta $x,p$ and $y,q$ for which a locality bound 
\eqn\gravbd{ |x-y|^{D-3}\geq G |p+q|}
is violated; here $G$ is a constant proportional to Newton's constant.  In the classical context, this would be roughly the criterion to produce a black hole.  A suggested generalization to the multimode case is outlined in \refs{\LQGST}.

Motivations to believe in such a bound include the following.  First, if one considers what locality means, it is commonly discussed in terms of local observables.  In the context of gravity, certain relational operators exist that reduce to local observables in certain circumstances\refs{\GMH}.  However, it is clear that there are profound obstacles to such a reduction to local observables for observations of modes that violate the locality bound \gravbd.  Alternatively, attempting to probe locality through high energy scattering encounters the same limitations\LQGST.
Secondly, if one believes that the Bekenstein-Hawking entropy accurately measures the number of degrees of freedom of a black hole, in accord with the holographic principle (see \refs{\Bous,\Bousrev}), this indicates that the local prediction of extensivity of the number of degrees of freedom with  the volume fails in situations where strong gravity is relevant.  Finally, local field theory leads one directly into the paradoxical results of information loss, which when combined with basic quantum principles, apparently lead to severe conflict with experiment\refs{\BPS,\SGinfo,\Astroinfo}.  The only apparent escape is some form of nonlocality, but it must be relevant only in restricted circumstances, and the connection of \gravbd\ with strong gravitational physics suggests it should be the appropriate limit.  

While these motivations exist, there is presently no direct {\it derivation} of a bound such as \gravbd.  Instead, the viewpoint of this paper is that this bound may be a fundamental new ingredient of quantum gravitational physics, which can't be truly derived from local quantum field theory plus perturbative gravity any more than that uncertainty principle can be derived from classical physics.  Thus the bound \gravbd\ will be taken as a {\it hypothesis}.

As a test for the relative roles of the gravitational nonlocality principle and the string uncertainty principle \genunc, one can explore properties of high energy scattering in string theory.  A summary of some of the relevant considerations appears in \LQGST, and further discussion appears below.  In short, in the high energy limit there is no evidence for activation of a bound \genunc, but there is evidence for a bound \gravbd.  In the context of string theory, it is also conceivable that there are new nonlocal physical effects at scales intermediate between \genunc\ and \gravbd, for a given high energy, as explained in \LQGST.   While this is an important avenue for further exploration, we will regard \gravbd\ as the ultimate limit.

Of course, a bound such as \gravbd\ would be only one constraint on fundamental physics,  and one expects others.  One is the absence of global charges, whose existence could equally well lead to paradoxical situations. One may be able to reason to this viewpoint by arguing that virtual effects of planckian physics are unitary for sectors with no global charge, and always violate global charge conservation.

If a hypothesized bound such as \gravbd\ holds, for consistency it is important to assess its relevance for explaining the appearance of the loophole in Hawking's argument for information loss.   A first requirement is extension of the bound \gravbd, which was motivated based on working about a flat background, to a more general statement about locality violation in a curved background.   A preliminary investigation of this problem appears in \refs{\SGnonloc}, which argued that for example in the background of a black hole, large kinematical invariants involving infalling modes and outgoing Hawking radiation (but {\it distinct} from the proposed role of ultraplanckian ``precursor" modes\refs{\tHooftRE\tHooftFN\tHooftFR\tHooftUN\VerlindeSG-\SchoutensST} of Hawking radiation) indicate the possibility of violating the analog of \gravbd, and thus suggest a rationale for relevance of the hypothesized nonlocal physics.   

This paper explores the question more carefully, and sharpens the argument.  Specifically, section two discusses generalities of the perturbation expansion for gravitational field.  Section three then discusses both the quantum and classical dynamics of a high energy collision in flat space.  There, the bound \gravbd\ is suggested to emerge in connection with breakdown of the gravitational perturbation expansion; the current proposal is that physics beyond the validity of this  approximation is unitary, but whatever unitarizes the physics is nonlocal.  There are different equivalent criteria for this breakdown, both quantum and classical.  Section four turns to quantum dynamics about black hole backgrounds.  Here, it is argued that there is an analogous breakdown of the gravitational perturbation expansion when attempting to simultaneously describe modes both inside the black hole and of late Hawking radiation.  This, again, apparently opens a window of opportunity for the hypothesized nonlocal physics to operate.

\newsec{Gravitational perturbation theory}

To set up the discussion, begin by considering the structure of the the perturbation expansion about a classical background $g_0$.  For greatest simplicity, consider gravity coupled to a scalar field $\phi$, which is assumed to have vanishing background value.  Gravitational amplitudes can be studied via the functional integral,
\eqn\gravfunc{\int_{g_i,\phi_i}^{g_f,\phi_f} \cald g \cald \phi e^{i(S[g] + S[g,\phi])}}
where $S[g]$ and $S[g,\phi]$ are the pure gravity and gravity-coupled matter actions, respectively. The amplitude \gravfunc\ can be convolved with definite initial and final states to obtain a corresponding transition amplitude.  For example, expand the fully fluctuating gravitational field,
\eqn\gravexp{g_{\mu\nu} = g_{0\mu\nu} + \sqrt {G_D} h_{\mu\nu}\ ,}
where $G_D$ is Newton's constant, and consider the amplitude for a transition between states $\Psi_i[h,\phi]$, $\Psi_f[h,\phi]$ that represent perturbations about this background:
\eqn\gravpert{\eqalign{\calA[\Psi_i,\Psi_f] = e^{iS[g_0]} \int_{\Psi_i}^{\Psi_f} &\cald h_{\mu\nu} \cald \phi \exp\Biggl\{ i \int d^Dx \sqrt{-g_0} (h\calL h + \sqrt{G_D}h^2\nabla^2 h + \cdots)\cr &+ i S[g_0,\phi] + i\sqrt{G_D}\int d^D x \sqrt{-g_0} h^{\mu\nu}T^\phi_{\mu\nu} + \cdots\Biggr\} \ .}}
In this expression $\calL$ is a second-order differential operator, cubic and higher terms are only schematically indicated, and $T^\phi_{\mu\nu}$ is the matter stress tensor computed using the background metric $g_0$.  

To evaluate the expression \gravpert\ one must confront some issues.  First, the diffeomorphism gauge symmetry must be fixed.  This symmetry manifests itself via arbitrariness in defining the decomposition \gravexp\ into background and perturbation; this results in the linearized gauge symmetry
\eqn\lindif{h_{\mu\nu}\rightarrow h_{\mu\nu} + \nabla^0_{(\mu} \xi_{\nu)}}
with parameter $\xi_\nu$.  This can in principle be treated by a standard gauge-fixing procedure, specifying a definite gauge and introducing a Fadeev-Popov determinant.  The resulting amplitudes also have infrared divergences, due to the massless graviton, but arguments have been given that these can be treated by Block-Nordsieck techniques, and moreover that IR safe results can be extracted\refs{\ACVthree}.  More importantly, the resulting perturbation expansion in $G_D$ is non-renormalizable: an infinite number of primitive divergences arise, and thus an infinite number of coupling constants must be specified.  Nonetheless, one can derive limited results from this perturbation expansion, viewing it as defining a low-energy effective field theory\refs{\Dono}, or even in high energy scattering, at sufficiently large impact parameter, where the eikonal approximation\refs{\MuSo\ACVone-\ACVtwo,\ACVthree,\KaOr\ACVfour-\GRW} connects to familiar results.  The latter are thus  apparently essentially insensitive to the ultimate ultraviolet physics of gravity.

The regime of perturbative gravitational physics is that where the perturbative dynamics in $h$ remains linear.  Once the higher-order terms in \gravpert\ become comparable to the quadratic and linear terms, the perturbative expansion fails.

In particular, one can give an approximate treatment of high energy, $E\gg M_p$, scattering that agrees well with a classical picture at large impact parameters.  Critical questions are when the perturbative treatment based on \gravpert\ fails, and what characterizes the physics that replaces it.

\newsec{High energy scattering in flat space}

More details can be supplied in the simple case of scattering of two very high energy $\phi$ particles in a flat background, $g_{0\mu\nu}=\eta_{\mu\nu}$.  We begin by summarizing some salient results of 
\refs{\MuSo\ACVone-\ACVtwo,\ACVthree,\KaOr\ACVfour-\GRW}, 
then discuss the connection with the classical picture, and finally comment on possible unitarization by nonlocal physics.  

The high energy problem  at center of mass energy $E=\sqrt s$ can be studied either in the context of pure gravity\refs{\ACVthree,\KaOr\ACVfour-\GRW} or in string theory\refs{\MuSo\ACVone-\ACVtwo,\GrMe}; comparison of results in the two was described in \refs{\LQGST}.   The Fourier transform on transverse momentum of the $\phi\phi\rightarrow\phi\phi$ (or two string to two string) element of the S-matrix  gives the amplitude in impact-parameter representation, which can be written as
\eqn\phasedef{S(b,E) = e^{2i \delta(b,E)}\ .}
Either in string theory or gravity, the leading-order eikonal amplitudes, resulting essentially from ladder diagrams with multiple graviton exchange between the two energetic particles, can be summed, with result
\eqn\treeph{\delta_0(b,E) \propto - {G_Ds \over \hbar b^{D-4}}\ }
(in $D=4$ the power is replaced by a log).  The leading eikonal phase $\delta_0$ is essentially the Fourier transform of the tree-level amplitude, $\calA_{tree}\sim G_Ds^2/t$,
\eqn\eiktree{\delta_0(|x_\perp|,E)\propto {1\over s} \int d^{D-2} k_\perp e^{-ik_\perp\cdot x_\perp} \calA_{tree}(s,t)\ ,}
modulo a numerical constant.

Gravitational corrections to this can be segregated into two categories.  First are the ``classical" corrections, also of order $\hbar^{-1}$; these essentially arise from diagrams where tree-level graviton scattering diagrams are exchanged between the two $\phi$ lines.  These result in a power series  in $(G_DE/b^{D-3})^2$.  The Schwarzschild radius corresponding to the center-of-mass energy $\sqrt s$ is 
\eqn\schdef{R_S(\sqrt s) \propto (G_D\sqrt s)^{1\over D-3}\ .}
The leading such term, from two loops, was computed (in $D=4$) in \ACVfour.  It has an IR divergent imaginary part, connected by unitarity to soft-graviton emission, and the real part is IR safe/convergent, and of the form
\eqn\deltatwo{{\it Re} \delta_2(b,E)= {2G_D^3 s^2\over \hbar b^2}\ .}
The second category of corrections are ``quantum," with powers $\hbar^0$ and higher.  The fact that the loop expansion includes terms that are an expansion in $R_S(\sqrt s)/b$ indicates that perturbation theory fails when impact parameters reach the Schwarzschild radius, along the lines described in section two.

In string theory, there is a third category of corrections, arising from intrinsically stringy effects.  One might have na\"\i vely expected that these begin to be important at $b\sim l_{st}^2 E$ (in accord with \genunc), but in fact the results of \refs{\ACVone,\ACVtwo} indicate that stringy corrections only become relevant at the scale $b\sim E^{2/(D-2)}$.  These ``diffractive string effects" 
apparently are well-explained in a semiclassical picture through
 local gravitational interactions between distant strings\LQGST:  graviton exchange can tidally excite higher modes of the individual strings.  By unitarity, this results in loss of amplitude from the original four-string process, but \ACVtwo\ argues that it is only some graying of the amplitude, and moreover is still a two-body process, which ultimately, again, has a loop expansion that breaks down at $b\sim R_S(\sqrt s)$ due to strong gravitational effects.  This picture is also supported, at the classical level, by arguments\LQGST\ that tidal string excitation doesn't spoil black hole formation.  Nonetheless, there could be new nonlocal effects in string theory at scales below $b\sim E^{2/(D-2)}$, and particularly possibly at scales\LQGST\ $b\roughly< E^{2/(D-1)}$.  Elucidation of string scattering in these regimes remains an interesting problem.

This quantum picture connects to a classical picture that has become increasingly precise.  (For some other discussion of the connection between the two, see \refs{\BaFi,\Hsu}.)  The classical metric of an ultrarelativistic particle is the Aichelburg-Sexl metric\refs{\AiSe}
\eqn\aisemet{ds^2= -dudv + dx^{i2} + \Phi(x^i)\delta(u) du^2\ .}
Here the flat spatial coordinates are $\{x^i, z\}$, null coordinates are $u=t-z$ and $v=t+z$, we define transverse radius $\rho^2=x^{i2}$ and unit $D$-sphere volume $\Omega_D$, and for a particle of energy $E$,
\eqn\aipot{\eqalign{ \Phi(\rho)& = -8G_DE\ln(\rho)\ ,\ D=4\cr 
\Phi(\rho)&= {16\pi G_D E\over (D-4)\Omega_{D-3} \rho^{D-4}}\ ,\ D>4\ .   }}
First consider test-particle motion in this metric.  It is straightforward to see that a test particle following a geodesic in the $-z$ direction with impact parameter $b$ is deflected by an angle given by
\eqn\angledef{\tan\theta(b) = {1\over 2} \partial_b \Phi(b)}
when it crosses crosses the shock at $u=0$.  This is a seemingly innocuous effect, and well-described perturbatively.  However, as soon as the energy of the test particle is taken into account, backreaction becomes relevant, and this leads to a breakdown of classical physics for sufficiently small impact parameter.  Specifically, in the center-of-mass frame, a closed trapped surface forms\refs{\EaGi} for $b\roughly<R_S(\sqrt s)$.  

This exemplifies the close connection between quantum and classical pictures.  Classically, at a critical impact parameter $b_c\sim R_S(\sqrt s)$ a black hole forms.  This happens when the tree-level center-of-mass scattering angle reaches a value $\theta\sim 1$.  Quantum-mechanically, this corresponds to a breakdown of the loop expansion; specifically, the series summing diagrams in the classical category diverges as the expansion parameter $R_S/b$ reaches $\calo(1)$.  A diagnostic for this in terms of the tree-level amplitude can be found from the relation
\eqn\classang{\theta_{cl} = -2\hbar {\partial \delta_0\over \partial L} = -{4 \hbar \over E} {\partial \delta_0 \over \partial b}\ ,}
together with the relation \eiktree\ between the one-graviton amplitude and the eikonal phase.
At the same time, the unitarity bound\refs{\Sold} for partial wave amplitudes $a_l(s)$,
\eqn\unitbd{|a_l(s)|\leq 1}
is saturated by the approximate eikonal amplitudes, as can be seen from expressions in \refs{\MuSo}.

A critical question driving at the heart of quantum gravitation is what physics takes over from the failed loop expansion in this regime; here, of course, nonperturbative effects should be relevant.  Hawking's original work \refs{\Hawkevap,\Hawkunc} proposed that this physics is not unitary, and instead produces a mixed state and superscattering operator.  Based on the arguments stated in the introduction, a more attractive picture is that the new physics unitarizes the S-matrix but has fundamental nonlocality on the scale $R_S$.  

A potential flaw in this picture is that, extending Hawking's analysis, it appears possible to develop a semiclassical picture of black hole formation which is well-described by local quantum field theory and semiclassical gravity on scales $\ll R_S(\sqrt s)$ (although even this semiclassical picture apparently requires accounting for some quantum effects\refs{\GiRy}).  This suggests that the divergent series giving the quantum amplitudes could be resummed and rewritten in terms of a perturbation expansion about the classical black hole geometry resulting from the collision.  While ultimately such resummed amplitudes should at least break down in the singular region, this could conceivably support a picture where unitarity itself is violated as originally advocated by Hawking.

Since such a picture (combined with basic quantum principles) is apparently in conflict with observation, it is important to try to understand where it could fail.  In order to do so, we turn to the closely related problem where one can apparently justify reexpressing the problem in terms of perturbations on the classical geometry corresponding to a black hole.

\newsec{Dynamics in black hole backgrounds}

Consider the problem of quantum evolution in the background geometry $g_0$ of a black hole of mass $M$, which for simplicity will be taken to be spherically symmetrical.  The most general form of Hawking's argument for information loss runs as follows.\foot{For more discussion see \refs{\SGnonloc}.}  One considers a set of slices in the background geometry, which are taken to asymptote to smooth spatial slices at infinity, which cut across both outgoing Hawking radiation and infalling matter, and which avoid regions of large curvature and are  ``as smooth as possible."   One then argues that the full quantum amplitudes on successive slices, given by \gravfunc, reduce, via the perturbative approximation \gravpert, to amplitudes for quantum matter fields on the classical background $g_0$; in other words, fluctuations of the metric must be {\it negligible}.  In this case, the state on the slices is well-described by local quantum field theory in a curved background, and locality ensures that the degrees of freedom inside the horizon do not influence the outgoing Hawking radiation.  More precisely, to describe outside observations, one should trace the state over internal degrees of freedom, producing a mixed-state density matrix.  Locality on scales $\ll R_S$ apparently ensures that this argument doesn't break down until the black hole reaches the Planck size, at which point energetics indicate that the information cannot escape except on a very long time scale that would lead to an unacceptable remnant scenario.  Thus in this picture the black hole disappears completely, leaving behind a mixed state.

\subsec{Perturbative quantum dynamics}

Next turn towards making this picture more precise.  Specifically, consider the case of a pre-existing black hole of mass $M$ into which falls a single quantum particle.  Ignoring for the moment the backreaction of the incident particle on the geometry, the slicing can be taken to be the nice slice construction of \refs{\LPSTU,\SGnonloc}, in which straight lines in Kruskal coordinates $U$ and $V$ (to be described further below) are matched onto a hyperbola at constant radius inside the horizon.  In the quantum state corresponding to the incident particle, the stress tensor $T^\phi$ in \gravpert\ will have two kinds of contributions: that of the incident particle, and that of the outgoing Hawking radiation.  Validity of the perturbation expansion about the background amounts to the statement that the fluctuations decouple, and correspondingly the perturbation $h_{\mu\nu}$ stays small.  A tree level indicator of the magnitude of such effects, as in the case of flat background, is the magnitude of the amplitude,
\eqn\treeM{\calA_{tree}= G_D\int \sqrt{-g} d^Dx \sqrt{-g} d^Dy T_{\mu\nu}(x) \Delta^{\mu\nu,\lambda\sigma}(x,y) T_{\lambda\sigma}(y)\ .}
Here $\Delta^{\mu\nu,\lambda\sigma}(x,y)$ is the Feynman propagator for the graviton, {\it i.e.} the inverse of the operator $\calL$ in \gravpert.  This must be defined in some gauge, but for conserved $T_{\mu\nu}$, the amplitude is invariant under \lindif.
Roughly, when $\calA_{tree}$ becomes large, this is an indication that self-gravitational effects of the quantum matter are becoming strong. In a flat background, a more precise criterion for breakdown of the perturbative expansion was that $\theta$ given by \classang\  becomes $\calo(1)$.  

Ref.~\SGnonloc\ sketched rough arguments that black hole kinematics can likewise produce a large $\calA_{tree}$ when considering interactions between infalling modes and late outgoing Hawking modes; basically, Kruskal kinematics demonstrates that the corresponding stress tensors of the modes experience a relative boost with a rapidity parameter that grows with the Schwarzschild time separation, producing a large invariant from the product of the stress tensors.  However, a complete calculation of the amplitude requires knowledge of the Feynman propagator $\Delta^{\mu\nu,\lambda\sigma}(x,y)$ in the Schwarzschild background.  Rough estimates in \SGnonloc\ suggested that the resulting expression could produce a large $\calA_{tree}$, but a more precise criterion is desired.

As part of this, one should also better understand the structure of the quantum stress tensor $T_{\mu\nu}$ that enters the expression \gravpert.  Notice that this is the stress tensor defined using the background $g_0$; the amplitude \treeM\ then summarizes the first contribution of its self-interaction.  The piece $T^1$ arising from the infalling mode is straightforwardly understood; one can for example treat the mode as a quantum wavepacket, producing a quantum stress tensor that can be expressed in the relevant frames defined by the slicing.  The piece $T^2$ corresponding to the Hawking radiation is somewhat more subtle, but still straightforwardly understood following the methods of \SGnonloc.  

Specifically, one can formally compute the expectation value $\langle T_{\mu\nu}\rangle$ for the Hawking state of the metric $g_0$; in two-dimensional models this can be done quite explicitly\refs{\CGHS,\GiNe}.  But moreover, one can compute the stress tensor for an individual Hawking mode, and then see how the sum of such expressions gives $\langle T_{\mu\nu}\rangle$.  To summarize arguments of \SGnonloc, one does this by finding solutions $v_{\omega l m}$ of the wave equation outside the horizon, with definite angular momenta, and whose radial wavefunctions have asymptotics 
\eqn\outasy{rR_{\omega l}e^{-i\omega t}\rightarrow  e^{-i\omega u} = \left(-{U\sqrt e\over 4M}\right)^{4iM\omega}\ }
in terms of Kruskal coordinate $U$ (see below) at the past horizon of the fully extended geometry.  One likewise needs corresponding solutions $\vhat_{\omega l m}^*$ inside the horizon, with asymptotic behavior 
\eqn\insideasy{ r{\hat R^*}_{\omega l}e^{i\omega t}\rightarrow  \left({U\sqrt e\over 4M}\right)^{4iM\omega}\ .}
Next, one defines combinations 
\eqn\vfd{\eqalign{v^1_{\omega lm} &= \left[v_{\omega lm}(U) \theta(-U) + \gamma_\omega {\hat v}^*_{\omega lm}(U)\theta(U)\right]/\sqrt{1-\gamma_\omega^2}\cr v^2_{\omega lm} &= \left[{\hat v}_{\omega lm}(U) \theta(U) + \gamma_\omega v^*_{\omega lm}(U)\theta(-U)\right]/\sqrt{1-\gamma_\omega^2}\ ,
}}
where
\eqn\gammafd{\gamma_\omega = e^{-4\pi M\omega}\ ;}
these are pure positive frequency in Kruskal time.  Wavepackets $v^1_K$, $v^2_K$ can be made by superposing the modes \vfd.  The conserved quantum stress tensor for a given outgoing quantum state of Hawking radiation is a superposition of terms of the form
\eqn\onestress{t_{K\mu\nu} = t_{\mu\nu}(v_K^{1*},v_K^{2*}) = \partial_{(\mu} v_K^{1*} \partial_{\nu)}v_K^{2*} - \hf g^0_{\mu\nu} \partial 
v_K^{1*}\cdot\partial v_K^{2*}}
for each excited Hawking mode $K$; the expectation value $\langle T_{\mu\nu}\rangle$ then comes from a thermal ensemble of such superpositions.

The following picture of the stress tensor for Hawking modes results.  A typical wavepacket $v_K$ of the Hawking radiation has width $\Delta U\sim U$ until it reaches a radius $\calo(M)$ from the horizon at $r=2M$.  Thus the mode and its inside partner $\vhat_K$ cannot be cleanly separated until this point.  Correspondingly the expression \onestress\ represents a  fluctuating stress tensor in this region.  However, when the mode reaches $r= 2M + \calo(M)$, it separates from its partner.  Moreover, the stress tensor takes on a classical form.  Outside the horizon, this is
\eqn\tout{t_{K\mu\nu} \propto t_{\mu\nu} (v_K^*,v_K)\ ;}
inside it is
\eqn\tin{t_{K\mu\nu} \propto t_{\mu\nu}(\vhat_K^*,\vhat_K)\ .}
Each of these are classical stress tensors corresponding to inside and outside propagating wavepackets; of course, by the time the inside packet reaches $r= 2M - \calo(M)$, it is in the strongly curved region near the singularity.  Thus, a picture of the stress tensor is that of modes that are born in pairs, at radii $r= 2M \pm \calo(M)$, when the width of the mode in $r$ is comparable to $M$; a rough picture is that of dripping from a faucet, where the size of the water droplet is comparable to the size of the original bulge of water.  Once the outside mode separates, it follows an essentially classical trajectory to infinity.  Note, however, that before it separates there can be backscattering of the mode that results in a flux also into the future horizon.

Thus, a fairly precise picture of the stress tensor in the background $g_0$ can be formulated.  One is however still limited by the lack of knowledge of the Feynman propagator $\Delta^{\mu\nu,\lambda\sigma}(x,y)$ in the Schwarzschild background.   

\subsec{Classical perturbations of Schwarzschild}

To circumvent this limitation, one can use the following observation.  In a given gauge, the classical perturbation in the metric resulting from a stress $T^1$ can be represented as
\eqn\Tpertmet{h^{1,\mu\nu}(x) = \sqrt{G_D} \int \sqrt{-g_0} d^Dy \Delta_R^{\mu\nu,\lambda\sigma}(x,y) T^1_{\lambda\sigma}(y)\ .}
Here the propagator is retarded, but only differs from the Feynman propagator on-shell.  Thus for off-shell graviton exchange, the size of \treeM\ can be estimated by combining the perturbation of the metric due to the incident particle $T^1$ with the stress tensor $T^2$ of the Hawking radiation.

While computation of $h^1$ is gauge dependent, the result \treeM\ should be gauge independent.  The gauge choice can be determined through specification of a definite slicing.  This allows us to compare the perturbed line element $ds^2$ with the original line element $ds_0^2$ to extract the perturbation $h_{\mu\nu}$.  In so doing, it is in particular important to begin with comparable intial data, and moreover to work with slices such that the asymptotic data differs at most by a true gauge transformation, that is, a diffeomorphism $\xi_\mu$ vanishing at infinity.

We will illustrate these points in a particularly simple case, replacing the stress tensor $T^1$ by that of a classical perturbation that corresponds to a spherically-symmetric massless perturbation that carries an energy $\delta M$ into the black hole along a given null trajectory, in dimension $D=4$.  The classical solution is that of Vaidya\refs{\Vaid},
\eqn\vsoln{ds^2 = -\left(1-{2M(v)\over r}\right) dv^2 + 2 dv dr + r^2 d\Omega_2^2\ ,}
where the mass function for incoming wave at advanced time $v=v_i$ is
\eqn\mfcn{M(v) = M + \delta M \theta(v-v_i)\ .}
While the perturbation appears small in these coordinates, this is not necessarily the correct measure.  In particular, one needs to choose some slicing, and compare the evolution on the slicing of the perturbed metric \vsoln\ and the original metric $ds_0^2$ with $M(v)\equiv M$.  This could for example be the nice slicing defined above; we would like a set of slices with definite unperturbed evolution in Schwarzschild time at infinity.  However, there is another subtlety, namely that the metrics $ds^2$ and $ds_0^2$ have different asymptotic behavior as $r\rightarrow\infty$, corresponding to their different masses.  This can be rectified by choosing a ``regulated" background metric $ds_R^2$ to also be of the form \vsoln, but with 
\eqn\regM{M(v) = M + \delta M \theta(v-v_R)}
for a very large  advanced time $v_R$.  The metrics $ds^2$ and $ds_R^2$ then have identical asymptotic behavior.  However, at times short as compared to $v_R$, the difference between the nice slicing of $ds_0^2$ and a nice slicing of $ds_R^2$ (asymptoting to a Schwarzschild time slicing at $r\rightarrow\infty$)  is small.  This means that we are justified in simply using a slicing defined by the original metric $ds_0^2$, up to a small error.

The easiest way to define such a slicing is to introduce the Kruskal coordinates.  Specifically, let 
\eqn\eddfink{v=t+r^*(r)\ ,\ u= t- r^*(r)}
where 
\eqn\tort{r^*(r) = r + 2M \ln \left({r-2M\over 2M}\right)\ .}
This can be used to bring $ds_0^2$ into the standard Schwarzschild form.  The Kruskal coordinates are defined by
\eqn\Kruskdef{U=-4M e^{-u/4M -1/2}\ ,\ V=4M e^{v/4M -1/2}\ ,}
and slicings (like that of \refs{\LPSTU,\SGnonloc}) can be given by functions of the form 
\eqn\slicedef{V = V_s(U)\ .}
In Kruskal coordinates, the background metric $ds_0^2$ becomes
\eqn\kruskmet{ds_0^2 = -{2M\over r} e^{1-{r\over 2M}} dU dV + r^2 d\Omega^2\ .}
For the perturbed metric $ds^2$ at $v>v_i$ one can likewise define quantities $r^{*\prime}$, $U'$, and $V'$ using $M'=M+\delta M$; the metric $ds^2$ takes the form \kruskmet\ in terms of the primed quantites.  However, we want to compare the two metrics on the family of slices, so we should write the metric $ds^2$ in terms of $U$, $V$.  This is accomplished by matching the two metrics along the interface $V=V_i$.  One can trivially match the $V$ coordinates, $V'=V$.  Next define the function $\rho(M,x)$ implicitly via
\eqn\rhodef{x=-8Me^{\rho/2M -1}(\rho - 2M)\ .}
Then matching the angular part of  metric along $V=V_i$ gives $U'(U)$, as the solution to
\eqn\rmatch{\rho(M,UV_i) = \rho(M',U'V_i)\ .}
Using this, the first term of the primed Kruskal metric can be converted into $U$ and $V$ coordinates.
The relation between the angular terms of the two metrics also immediately follows:
\eqn\raddefs{ r(U,V) = \rho(M,UV)\ ,\ r'(U,V) = \rho(M', U'(U)V)\ .}

With these definitions, one can now find $h_{\mu\nu}$, from
\eqn\hdef{h_{\mu\nu} dx^\mu dx^\nu = ds^2 -ds_0^2}
evaluated as above in the $U,V$ coordinates.  

The large contribution to \treeM\ arises as follows.  First, recall that the stress tensor of Hawking modes in the background metric can be thought of as describing outgoing particles produced at 
\eqn\prodr{r(U,V)=2M +\calo(M)\ .}
However, in the perturbed metric $ds^2$, these curves dip inside the horizon, which lies at $U'(U)=0$, for sufficiently late $V$.  Thus, the perturbed metric has a large effect on the Hawking stress tensor; the outgoing Hawking particles are now pulled into the (perturbed) singularity, which lies along the curve $r'(U,V)=0$.  

A rough estimate of when this effect becomes important can be made by computing the perturbation of the position of the horizon, given by evaluating $\delta U' = U'-U$ at $U=0$, to linear order in $\delta M$.  One readily finds 
\eqn\deltau{\delta U' = {16M\over V_i} \delta M\ .}
The line $r=2M +\calo(M)$ thus gets pulled into the new horizon for times $V_r$ such that 
\eqn\retV{ {V_r\over V_i} \delta M \sim M\ ,}
or for Schwarzschild time differentials 
\eqn\rettime{t_r-t_i \sim4 M \ln (M/\delta M)\ }
comparable to the more na\"\i ve estimates of \SGnonloc.

Before turning to the proposed quantum interpretation of this result, we digress with a brief discussion of nice slices.  As mentioned, one defines the slicing using the Kruskal coordinates of the background metric (or more precisely its regulated version $ds_R^2$, which represents a small difference for the times that concern us).  Recall that these slices should (for making Hawking's argument in an asymptotically flat metric, such as we are considering) asymptote to slices of constant Schwarzschild time $t$ at $r\rightarrow\infty$, and cut across both  outgoing Hawking modes and infalling matter.  One choice is that of \refs{\LPSTU,\SGnonloc}, in which the $T=0$ slice is defined to be the horizontal curve $V+U=const$, matched to a hyperbola $r=r_c$ inside the horizon.  Subsequent slices are found by pushing forward by a Schwarzschild time translation.  Coordinates along the slice can be taken to be the radius $r$ for $r\geq r_c$, and the Schwarzschild time (which is proportional to proper distance along the slice) along the hyperbolic portion.  This family of slices can likewise be trivially mapped into slices in the perturbed geometry $ds^2$.  Once the slice intersects $r'=r_c$, one should include a portion of that hyperbola.  Note that there are then kinks in the nice slices where $r=r'=r_c$, and where the straight section intersects $r'=r_c$.  

There is one other somewhat pathological feature of nice slices: along the hyperbolic section, the timelike part of the metric vanishes, since there must be zero elapsed proper time for infinite coordinate time.  One could regulate this by taking as slices a sequence of hyperbolae (and corresponding slightly more complicated matching conditions) whose radii shrinks as nice-slice time $T$ increases, say as some function $r_c=r_c(T)$.  However, since by definition the slices must avoid the strong curvature region at $r=0$, they must asymptote to a finite value $r_c(\infty)$, and correspondingly the timelike part of the metric asymptotes to zero.  Notice that this introduces pathologies as well into the standard form of the constraint equations $G^0_0 = 8\pi G_NT^0_0$, $G^0_1 = 8\pi G_NT^0_1$.  In particular, the momentum $T^0_1$ becomes divergent.  

\subsec{Fate of the perturbative expansion}

To summarize so far, we have set up a perturbative expansion \gravpert\ for quantum perturbations of a Schwarzschild metric.  An essential question is whether this perturbation expansion stays well controlled for small fluctuations of the infalling matter, in the sense discussed in section two.  A test of this is the relative size of the amplitude \treeM\ and subleading perturbations.  While our understanding of the stress tensors in these expressions, including the Hawking radiation, is fairly good, we don't know the Feynman propagator in the Schwarzschild background.  However, one can estimate the amplitude \treeM\ using the fact that part of it arises from the retarded field caused by the perturbation $T^1_{\mu\nu}$ of an infalling quantum.  According to this estimate, it gets large when comparing such a perturbation, and Hawking modes emitted at a later time given by \rettime.  This suggests that the full gravitational loop expansion breaks down, and that one cannot treat the problem in terms of small perturbations of the background metric $g_0$.  Notice, furthermore, that the infalling quantum could be the backscattered part of an earlier Hawking mode.   The large backreaction potentially leads to large interactions between fluctuations at different times.

As with a flat background, the precise condition is not largeness of the tree amplitude, but its size relative to the subleading terms.  Again, these can't be precisely calculated, but drawing from that analogy, breakdown of the ``classical" part of the loop expansion occurs when the classical scattering deflection reaches order unity.  On the classical side, it is apparent where that occurs in the black hole background.  Once the condition \retV\ or \rettime\ is satisfied, the corresponding Hawking radiation falls into the singularity; this is an indicator of the breakdown.\foot{Note another potential source of semiclassical breakdown is the divergent momentum mentioned in the preceding subsection.  However, at the classical level, this doesn't seem to produce a big effect on the outgoing Hawking modes, beyond those already discussed.}

This strong classical effect should of course not be the full story, but rather is taken to be indicative that the loop expansion has failed and must be replaced by some other dynamics, as in the flat background.  One may suggest that this expansion could be resummed to give an again semiclassical description.  However, it is far from clear that this can be done.  In short, our straightforward attempt to argue that the loop expansion can be used and reduces to local quantum field theory on slices in the background metric of the black hole has failed.

The options seem to be the following.  One is that the loop expansion can be resummed, and cast into a form described by local physics.  Or, perhaps the full amplitude is computable in terms of some other local physics.  Either of these cases would support a picture where the information is lost to the interior of the black hole, resulting in Hawking's claimed violation of unitarity, and conflict with observation.  The third alternative is that the loop expansion breaks down and is replaced by some fundamentally nonlocal but unitary gravitational physics, in accord with the nonlocality principle.  While novel, this alternative suggests a picture in accord with observation; moreover, the proposed nonlocality only arises in circumstances of extreme kinematics, and thus would apparently not have been observed elsewhere.  

The arguments given above share some common elements with discussions of the ``S-matrix Ansatz" pursued by 't Hooft and others\refs{\tHooftTQ\KiemIY-\ArcioniWC}.  However, there are important differences.  Notably, a role has not been assumed for ultraplanckian modes that arise from tracing Hawking particles back to near the horizon; as argued in \SGnonloc, these modes are not expected to have interactions with other modes.  Moreover, the relevant dynamics of gravity is much more than the shifts described in those references; in the flat space case these shifts are only a leading indicator of black hole formation and breakdown of the perturbative expansion.  Finally, once one encounters a regime where the gravitational perturbation expansion apparently fails, one still needs a rationale for nonlocal physics; the present paper thus makes the postulate that the nonperturbative dynamics of gravity is intrinsically nonlocal.

As a final note, if the correct theory of quantum gravity is string theory, combined string and gravitational effects could conceivably lead to breakdown of the effective theory even earlier.  While there is no concrete evidence for this, the discussion of high-energy scattering \LQGST\ suggests that the earliest one might expect this to occur would be when tidal string deformation effects become important, analogous to the dynamics at impact parameters $b\roughly< E^{2/D-1}$ in a collision in flat space.

\newsec{Conclusions}

In order to avoid the black hole information paradox, it appears that a principle of local quantum field theory must be sacrificed.  Violation of unitary evolution appears inconsistent with observation\BPS.  The only obvious alternative is sacrifice of locality.  However, if physics is fundamentally nonlocal, it certainly {\it appears} well-described by local quantum field theory in directly accessible circumstances.  There should thus be a correspondence between the hypothesized nonlocal physics and local quantum field theory (including perturbative gravity) in a definite regime.  A straightforward attempt to parameterize that regime  gives the bound \gravbd\ as the dividing line between the local and nonlocal domains.

The limits of local physics can in principle be probed either by study of the gravity-coupled analogs of local observables\GMH\ or through high energy scattering\refs{\ACVone,\ACVtwo,\BaFi,\LQGST}.  Gedanken experiments in either case suggest that the expression \gravbd\ indeed serves as a bound on a local description.  

However, consistency of such a picture requires an understanding of why local quantum field theory doesn't suffice to describe states on spatial slicings that intersect both matter that has fallen into a black hole, and late-time Hawking radiation; if such a description did exist, it would both suggest nonunitary evolution of black holes\Hawkunc\ and remove any rationale for the hypothesized nonlocality.  This paper has investigated the gravitational perturbation theory on such slices, and found indications that a na\"\i ve analysis fails.  In short, the condition that matter field perturbations not drive large gravitational perturbations, which holds for example in low-energy scattering in a flat background, is apparently violated on these slices.  Evidence for this is seen by studying the retarded gravitational field of a matter-field perturbation, and its effect on outgoing Hawking modes.   A fully consistent study of the gravitational perturbation theory would involve instead the Feynman propagator for gravitons.  Due to the half-retarded, half-advanced nature of the propagator, this appears even more problematic, as its definition apparently requires deeper understanding of propagation inside the black hole.  Thus, while one cannot yet state conclusively that no perturbative analysis exists, a straightforward attempt to derive one fails. If it is true that no such perturbative analysis exists, similar arguments likely have implications in other quarters, particularly for certain aspects of cosmology.

An iron-clad proof that {\it no} local perturbative description of physics exists on such slicings of black holes or other geometries
 would be even more satisfying, but may be impossible to give.  However, one should consider that the absence of such a proof is not necessarily a fundamental shortcoming.  Indeed, returning to the analogy of the development of quantum mechanics, imagine the plight of a classical physicist trying to take the same approach.  Our bewildered classical physicist would know that there is a problem -- classical physics predicts the instability of the hydrogen atom.  He could then attempt to give a very careful treatment of the classical evolution of the  atom, and try to infer exactly where, just before the electron's orbit becomes singular, classical physics fails.  But such an approach would never reveal the correct quantum mechanical answer.  Rather, in a quantum framework it is just true that a classical description is no longer valid in a certain regime, roughly characterized by the uncertainty principle, and this is what prevents the disasterous classical instability.  Likewise, in gravitational physics it could be that a local quantum description is simply no longer correct in a certain regime, but one may not be able to prove the existence of the boundary of that regime within that local quantum framework, any more than a classical physicist could prove the existence of quantum mechanics.

\bigskip\bigskip\centerline{{\bf Acknowledgments}}\nobreak

The author wishes to thank 
J. Hartle,  G. Horowitz, J. Polchinski, R. Sugar, and G. Veneziano for important discussions.
This work  was supported in part by the Department of Energy under Contract DE-FG02-91ER40618.


\listrefs
\end

\end